\documentclass[letterpaper]{JHEP3}
\usepackage{amsmath}
\usepackage{dsfont}
\usepackage{bm}

\newcommand{\eq}{\begin{equation}}
\newcommand{\eeq}{\end{equation}}

\title{ Planar AdS black holes in Lovelock gravity  with a nonminimal scalar field}

\author{Mois\'es Bravo Gaete\\
Instituto de Matem\'atica y F\'{\i}sica, Universidad de Talca,\\
Casilla 747, Talca, Chile.\\
E-mail: \email{mbravog-at-inst-mat.utalca.cl}}

\author{Mokhtar Hassa\"{\i}ne\\
Instituto de Matem\'atica y F\'{\i}sica, Universidad de Talca,\\
Casilla 747, Talca, Chile.\\
E-mail: \email{hassaine-at-inst-mat.utalca.cl}}

\abstract{In arbitrary dimension $D$, we consider a self-interacting
scalar field nonminimally coupled with a gravity theory given by a
particular Lovelock action indexed by an integer $k$. To be more
precise, the coefficients appearing in the Lovelock expansion are
fixed by requiring the theory to have a unique AdS vacuum with a
fixed value of the cosmological constant. This yields to
$k=1,2,\cdots,[\frac{D-1}{2}]$ inequivalent possible gravity
theories; here the case $k=1$ corresponds to the standard
Einstein-Hilbert Lagrangian. For each par $(D,k)$, we derive two
classes of AdS black hole solutions with planar event horizon
topology  for particular values of the nonminimal coupling
parameter. The first family of solutions depends on a unique
constant and is valid only for $k\geq 2$. In fact, its GR
counterpart $k=1$ reduces to the pure AdS metric with a vanishing
scalar field. The second family of solutions involves two
independent constants and corresponds to a stealth black hole
configuration; that is a nontrivial scalar field together with a
black hole metric such that both side of the Einstein equations
(gravity and matter parts) vanishes identically. In this case, the
standard GR case $k=1$ reduces to the Schwarzschild-AdS-Tangherlini
black hole metric with a trivial scalar field. We show that
the two-parametric stealth solution defined in $D$ dimension
can be promoted to the uniparametric black hole solution  in $(D+1)$  dimension
through a {\it Kaluza-Klein oxidation}. In both cases, the
existence of these solutions is strongly inherent of the presence of
the higher order curvature terms $k\geq 2$ of the Lovelock gravity.
We also establish that these solutions emerge from a stealth
configuration defined on the pure AdS metric through a Kerr-Schild
transformation. Finally, in the last part, we include multiple exact
$(D-1)-$forms homogenously distributed and coupled to the scalar
field. For a specific coupling, we obtain black hole solutions for
arbitrary value of the nonminimal coupling parameter generalizing
those obtained in the pure scalar field case.}

\begin{document}

\section{Introduction}
Since the advent of string theory, the interests on
higher-dimensional physics have grown up in the last decades, and
particulary concerning the higher-dimensional General Relativity
(GR). This latter is realized in the lowest order of the Regge slope
expansion of strings. It is believed that string corrections to the
standard higher-dimensional Einstein-Hilbert action may be source of
inconsistencies. Indeed, these corrections being given by some
higher powers of the curvature may yield to field equations of order
greater than two or may introduce ghosts. Moreover, as shown in
\cite{Zw} and \cite{Zumino}, in order to the graviton amplitude to
be ghost-free a special combination of quadratic corrections which
is nothing but the Gauss-Bonnet expression is required. One of the
interesting features of the Gauss-Bonnet Lagrangian lies in the fact
that its variation yields second-order field equations for the
metric in spite of the presence of quadratic terms in the curvature.
The Einstein-Hilbert-Gauss-Bonnet gravity piece is part of a more
general gravity theory build out of the same principles as GR.
Indeed, two of the main fundamental assumptions in GR are the
requirements of general covariance and the fact that the field
equations for the metric to be at most of second order. In view of
this, it is natural to describe the spacetime geometry in three or
four dimensions by the standard Einstein-Hilbert action (with
eventually a cosmological constant term) while for dimensions
greater than four a more general theory can be used. This fact has
been first noticed by Lanczos in five dimensions \cite{LAN} and then
generalized in higher dimension $D$ by Lovelock \cite{LOV}. The
resulting action is the so-called Lovelock gravity action which is a
$D-$form constructed out of the vielbein, the spin connection and
their exterior derivative. By construction, the Lovelock Lagrangian
which contains higher powers of the curvatures remains invariant
under local Lorentz transformations. In odd dimension, this gauge
symmetry can be extended to a local anti de Sitter (AdS) or
Poincar\'e symmetry through a particular choice of the coefficients
appearing in the Lovelock expansion. In both cases, the resulting
Lagrangian is a Chern-Simons form since its exterior derivative is
an invariant homogeneous polynomial in the AdS or Poincar\'e
curvatures, and their supersymmetric extensions are also known; see
\cite{Zanelli:2005sa} for a good review on Chern-Simons
(super)gravity. The Lovelock gravity or its Chern-Simons particular
case have been shown to possess (topological) AdS black hole
solutions with interesting thermodynamical properties
\cite{Banados:1993ur,Cai:1998vy,Crisostomo:2000bb,Aros:2000ij}
generalizing those obtained in the Einstein-Gauss-Bonnet case
\cite{Boulware:1985wk,Cai:2001dz}; for good reviews on
Einstein-Gauss-Bonnet black holes, see e.g. \cite{Charmousis:2008kc,Garraffo:2008hu,Camanho:2011rj}.

In the present work, we will consider a gravity action given by a
particular Lovelock Lagrangian with fixed coefficients such that the
resulting theory has a unique anti-de Sitter vacuum with a fixed
cosmological constant while our matter action will be concerned with
a nonminimal self-interacting scalar field. For this Lovelock
gravity model with nonminimal scalar field source, we will look for
topological black hole solutions with planar event horizon topology.
Note that  the first examples of topological black holes in GR
without source were discussed in \cite{Lemos}. The reasons of
considering such a matter source are multiples. Firstly, the ideas
behind the anti-de Sitter/Conformal Field Theory (AdS/CFT)
correspondence \cite{Maldacena:1997re} have been recently extended
to non-relativistic physics particularly to gain a better
understanding of some unconventional superconductors
\cite{Horowitz:2010gk,Hartnoll:2008kx,Herzog:2009xv}. In this
context, black holes with scalar hair at low temperature which must
disappears at low temperature are important in order to reproduce
the correct behavior of the superconductor phase diagram.  It is
well-known now from the BBMB solution (solution of the Einstein
equations with a conformal source given by a scalar field) in four
dimensions \cite{Bekenstein:1974sf,Bocharova:1970} that scalar
fields  nonminimally coupled can be useful to escape standard
no-hair theorems \cite{Bekenstein:1998aw}. Note that the BBMB
solution has been extended in presence of a cosmological constant
with a potential term in four dimensions and for the conformal
nonminimal coupling parameter $\xi=1/6$,
 \cite{Martinez:2002ru,Martinez:2005di,Anabalon:2012tu,Anabalon:2012ta}.  In fact, scalar fields nonminimally coupled to curved spacetimes play an
important role in different branches of physics and are also of
interest for pure mathematical proposals (as for example for the
Yamabe problem). The introduction of nonminimal couplings in spite
of complicating the calculations may be of extremely relevance for
many problems. For example, the nonminimal couplings are generated
by quantum corrections even if they are absent in the classical
action \cite{Linde:1982zj}, and they are required in order to
renormalize the theory or at least to enhance their
renormalizability properties
\cite{Callan:1970ze,Freedman:1974gs,Freedman:1974ze}. In
cosmological context, it has been argued that in most of the
inflationary scenarios with scalar fields, the presence of the
nonminimal coupling is unavoidable and its correct value depends on
the gravity and the scalar field models adopted
\cite{Faraoni:1996rf,Faraoni:1997fn}. Secondly, we have already
consider such matter source in the case of a particular combination
of the Einstein-Hilbert-Gauss-Bonnet gravity action and establish
the existence of some black hole configurations for particular
values of the nonminimal coupling parameter \cite{Gaete:2013ixa}.
The present work is then the natural extension of the work done in
\cite{Gaete:2013ixa} in order to reinforce our conviction that these
black hole solutions are strongly inherent to the presence of the
higher-order curvature terms.

The plan of the paper is organized as follows. In the next section,
we present the model of a scalar field nonminimally coupled with a
gravity action given by a particular Lovelock Lagrangian. After
deriving the field equations, we will present two classes of
topological black hole solutions with planar base manifold. In Sec.
$3$, we will add to the starting action exact $(D-1)-$forms and
obtain a more general class of black hole solutions generalizing
those obtained in the pure scalar field case. Finally, the last
section is devoted to our conclusions, comments and further works.
Two Appendices are also added. In the first one, we show that these
black hole solutions can be constructed from a stealth configuration
on the pure AdS metric through a Kerr-Schild transformation. In the
second Appendix, we establish that the existence of these solutions
is inherent to the higher-order  curvature terms $k\geq 2$ of the
Lovelock Lagrangian and they can not be promoted to black hole
solutions in the standard GR case $k=1$.

\section{Planar AdS black holes for a particular Lovelock  gravity with a nonminimal scalar field}
We start with a generalization of the Einstein-Hilbert  gravity
action in arbitrary dimension $D$ yielding at most to second-order
field equations for the metric and known as the Lovelock Lagrangian.
This latter is a $D-$form constructed with the vielbein $e^a$, the
spin connection $\omega^{ab}$, and their exterior derivatives
without using the Hodge dual. The Lovelock action is a polynomial of
degree $[D/2]$ (where $[x]$ denotes the integer part of $x$) in the
curvature two-form, $R^{ab} = d\,\omega^{ab} + \omega^{a}_{\;c}
\wedge \omega^{cb}$ as
\begin{subequations}
\begin{eqnarray}
&&\int \sum_{p=0}^{[D/2]}\alpha_p~ L^{(p)},\\
&& L^{(p)}=\epsilon_{a_1\cdots a_d} R^{a_1a_2}\cdots
R^{a_{2p-1}a_{2p}}e^{a_{2p+1}}\cdots  e^{a_d},
\end{eqnarray}
\end{subequations}
where the $\alpha_p$ are arbitrary dimensionful coupling constants
and where wedge products between forms are understood. Here
$L^{(0)}$ and $L^{(1)}$ are proportional respectively to the
cosmological term and the Einstein-Hilbert Lagrangian. Now, as shown
in Ref. \cite{Crisostomo:2000bb}, requiring the Lovelock  action to
have a unique AdS vacuum with a unique cosmological constant, fixes
the $\alpha_p$  yielding to a series of actions indexed by an
integer $k$, and given by
\begin{eqnarray}
I_k=-\frac{1}{2k(D-3)!}\int \sum_{p=0}^{k}\frac{C^k_p}{(D-2p)}~
L^{(p)},\qquad\qquad 1\leq k\leq \Big[\frac{D-1}{2}\Big], \label{Ik}
\end{eqnarray}
where $C^k_p$ corresponds to the combinatorial factor. The global
factor in front of the integral is chosen such that the gravity
action (\ref{Ik}) can be re-written in the standard fashion as
\begin{eqnarray}
I_k=\frac{1}2\int d^{D}x\,\sqrt{-g} \Big[&&R+\frac{(D-1)(D-2)}{k}+\frac{(k-1)}{2(D-3)(D-4)}{\cal L}_{GB}+\nonumber\\
&& \frac{(k-1)(k-2)}{3!(D-3)(D-4)(D-5)(D-6)}{\cal L}_{(3)}+\cdots
\Big] , \label{Ik2}
\end{eqnarray}
where ${\cal L}_{GB}=R^{2}-4\,R_{\mu \nu}R^{\mu
\nu}+R_{\alpha\beta\mu\nu}R^{\alpha\beta\mu\nu}$ stands for the
Gauss-Bonnet Lagrangian, and  ${\cal L}_{(3)}$ is given by
\begin{eqnarray*}
{\cal L}_{(3)}&=&R^3   -12RR_{\mu \nu } R^{\mu \nu } + 16\,R_{\mu
\nu }R^{\mu }_{\phantom{\mu} \rho }R^{\nu \rho } + 24 R_{\mu \nu
}R_{\rho \sigma }R^{\mu \rho \nu \sigma }+ 3RR_{\mu \nu \rho \sigma
} R^{\mu \nu \rho \sigma }
\nonumber \\
&&-24R_{\mu \nu }R^{\mu} _{\phantom{\mu} \rho \sigma \kappa } R^{\nu
\rho \sigma \kappa  }+ 4 R_{\mu \nu \rho \sigma }R^{\mu \nu \eta
\zeta } R^{\rho \sigma }_{\phantom{\rho \sigma} \eta \zeta }-8R_{\mu
\rho \nu \sigma } R^{\mu \phantom{\eta} \nu \phantom{\zeta}
}_{\phantom{\mu} \eta \phantom{\nu} \zeta } R^{\rho \eta \sigma
\zeta }.
\end{eqnarray*}
Note
that in odd dimension $D=2n-1$ and for $k=n-1$, the corresponding
action $I_{n-1}$ is a Chern-Simons action, that is a $(2n-1)-$form
whose exterior derivative can be written as the contraction of an
invariant tensor with the wedge product of $n$ curvatures two-form.
In even dimension $D=2n$, the maximal value of $k$ is $n-1$, and in
this case the resulting gravity action has a Born-Infeld like
structure since it can be written as the Pfaffian of the $2-$form
$\bar{R}^{ab}=R^{ab}+e^ae^b$.  The gravity theories $I_k$ have been
shown to possess black hole solutions with interesting features, in
particular concerning their thermodynamics properties, see
\cite{Crisostomo:2000bb} and \cite{Aros:2000ij}. In what follows, we
will consider a scalar field nonminimally coupled together with the
gravity actions given by $I_k$, (\ref{Ik}). More precisely, we will
consider the following action for any integer $k\geq 2$\footnote{The
standard GR case $k=1$ will be discussed in the Appendix.}
\begin{align}
S_k=&I_k-\int d^{D}x\,\sqrt{-g} \Big[
\frac{1}{2}\partial_{\mu}\Phi\partial^{\mu}\Phi+\frac{\xi}2 R\Phi^2
+U(\Phi)\Big], \label{Sk}
\end{align}
The field equations read
\begin{subequations}
\label{eqsmotionk}
\begin{eqnarray}
&&{\cal G}^{(k)}_{\mu\nu}=T_{\mu\nu},\\
&&\Box \Phi = \xi R \Phi+\frac{d U}{d\Phi},
\end{eqnarray}
\end{subequations}
where ${\cal G}^{(k)}_{\mu\nu}$ is the gravity tensor associated to
the variation of the action $I_k$ (\ref{Ik}),
\begin{eqnarray*}
{\cal G}^{(k)}_{\mu\nu}=&&G_{\mu\nu}-\frac{(D-1)(D-2)}{2k}g_{\mu\nu}+\frac{(k-1)}{2(D-3)(D-4)}K_{\mu\nu}+\nonumber\\
&&\frac{(k-1)(k-2)}{3!(D-3)(D-4)(D-5)(D-6)}S_{\mu\nu}+\cdots
\end{eqnarray*}
where $K_{\mu\nu}$ is the Gauss-Bonnet tensor
$$
K_{\mu\nu}=
2\big(RR_{\mu\nu}-2R_{\mu\rho}R^{\rho}_{\phantom{\rho}\nu}-2R^{\rho\sigma}R_{\mu\rho\nu\sigma}
+R_{\mu}^{\phantom{\mu}\rho\sigma\gamma}R_{\nu\rho\sigma\gamma}\big)
-\frac{1}{2}\, g_{\mu\nu}\mathcal{L}_{GB}
$$
and $S_{\mu\nu}$ arises from the variation of ${\cal L}_{(3)}$,
\begin{eqnarray*}
S_{\mu\nu}&=&\,3\Big(R^2 R_{\mu \nu } - 4 R R_{\rho  \mu } R^{\rho}
_{\phantom{\rho} \nu } - 4R^{\rho \sigma }R_{\rho \sigma }R_{\mu \nu
} +8 R^{\rho \sigma }R_{\rho  \mu }R_{\sigma  \nu } - 4 R R^{\rho
\sigma }R_{\rho \mu  \sigma  \nu }
\nonumber \\
&&  +8 R^{\rho \kappa }R^\sigma _{\phantom{\sigma }\kappa }R_{\rho
\mu  \sigma  \nu } -16 R^{\rho \sigma }R^\kappa _{\phantom{\kappa }
(\mu }R_{ |\kappa \sigma \rho |  \nu ) } + 2 R R^{\rho \sigma \kappa
}_{\phantom{\rho \sigma \kappa }\mu  }R_{\rho \sigma \kappa  \nu }
+R_{\mu \nu }R^{\rho \sigma \kappa \eta } R_{\rho \sigma \kappa \eta
}
\nonumber \\
&&- 8 R^\rho _{\phantom{\rho }(\mu }R^{\sigma \kappa \eta
}_{\phantom{\sigma \kappa \eta } |\rho | }R_{|\sigma \kappa \eta |
\nu ) } - 4 R^{\rho \sigma }R^{\kappa \eta }_{\phantom{\kappa \eta }
\rho \mu }R_{\kappa \eta  \sigma  \nu }+8R_{\rho \sigma }R^{\rho
\kappa  \sigma  \eta }R_{\kappa  \mu  \eta  \nu } - 8 R_{\rho \sigma
}R^{\rho  \kappa \eta }_{\phantom{\rho  \kappa \eta }\mu }R^\sigma
_{\phantom{\sigma } \kappa \eta  \nu }
\nonumber \\
&&+4 R^{\rho \sigma  \kappa \eta }R_{\rho \sigma  \zeta  \mu
}R_{\kappa \eta  \phantom{\zeta } \nu }^{\phantom{\kappa \eta }\zeta
} - 8 R^{\rho  \kappa  \sigma  \eta }R^\zeta _{\phantom{\zeta }\rho
\sigma  \mu }R_{\zeta  \kappa \eta  \nu }- 4R^{\rho \sigma \kappa
}_{\phantom{\rho \sigma \kappa } \eta  } R_{\rho \sigma \kappa \zeta
}R^{\eta  \phantom{\mu }\zeta }_{\phantom{\eta } \mu \phantom{\zeta
} \nu }\Big)-  \frac12 g_{\mu \nu }{\cal L}_{(3)}.
\end{eqnarray*}
In the
matter part of the equations (\ref{eqsmotionk}), $T_{\mu\nu}$
represents the energy-momentum tensor of the scalar field given by
\begin{eqnarray}
\label{tmunusf} T_{\mu \nu}=&&\partial_{\mu}\Phi\partial_{\nu}
\Phi-g_{\mu
\nu}\Big(\frac{1}{2}\,\partial_{\sigma}\Phi\partial^{\sigma}\Phi+U(\Phi)\Big)+\xi\left(g_{\mu
\nu}\Box-\nabla_{\mu}\nabla_{\nu}+G_{\mu\nu}\right)\Phi^{2},
\end{eqnarray}
where the potential $U(\Phi)$  is given by a mass term
\begin{equation}\label{potstealth}
U(\Phi)=\frac{8\,\xi\,D\,(D-1)}{(1-4\xi)^{2}}\,(\xi-\xi_{D})(\xi-\xi_{D+1}){\Phi}^{2},
\end{equation}
where $\xi_{D}$ denotes the conformal coupling in $D$ dimensions
\begin{equation}\label{conformal}
\xi_{D}=\frac{D-2}{4(D-1)}.
\end{equation}
The choice of such potential will be justified in the Appendix.
Being a mass term and because of the presence of the term $\xi
R\Phi^2$ in the action, one can define an effective mass
$m_{\mbox{\tiny{eff}}}^2$ in the case of solution of constant
curvature $R=\mbox{constant}$ by
\begin{eqnarray}\label{effmass}
m_{\mbox{\tiny{eff}}}^2=\xi
R+\frac{16\,\xi\,D\,(D-1)}{(1-4\xi)^{2}}\,(\xi-\xi_{D})(\xi-\xi_{D+1}).
\end{eqnarray}

As for the Einstein-Gauss-Bonnet case $k=2$ \cite{Gaete:2013ixa}, we
will obtain the same two classes of black hole solutions for generic
value of $k\geq 2$. More precisely, for each par $(D,k)$ with $D\geq
5$ and $k\geq 2$ , we will derive two classes of AdS black hole
solutions with planar event horizon topology for specific values of
the nonminimal coupling parameter $\xi$.

\subsection{Planar AdS  black hole solutions}

For $k\geq 2$, an AdS black hole solution is obtained provided that
the nonminimal parameter $\xi$ takes the following form
\begin{eqnarray}
\label{xikbh}
\xi_{k,D}^{\tiny{\mbox{b.h}}}=\frac{(D-2)(k-1)}{4\Big[(D-1)k-(D-2)\Big]},
\end{eqnarray}
which in turn implies that the potential (\ref{potstealth}) becomes
\begin{eqnarray}
U_{k,D}^{\tiny{\mbox{b.h}}}(\Phi)=\frac{(k-1)(D-2)^2(D-2+k)}{8k^2\Big[(D-1)k-(D-2)\Big]}\Phi^2.
\label{potbh}
\end{eqnarray}
In this case, the metric solution and the scalar field are given by
\begin{eqnarray}
\label{solbhkd}
&&ds^{2}=- F_{k,D}^{\tiny{\mbox{b.h}}}(r)dt^{2}+\frac{dr^2}{ F_{k,D}^{\tiny{\mbox{b.h}}}(r)}+{r^{2}}d\vec{x}_{D-2}^2,\nonumber\\
&& F_{k,D}^{\tiny{\mbox{b.h}}}(r)=r^2-\frac{M}{r^{\frac{D-2(k+1)}{k}}},\\
\nonumber\\
&&
\Phi_{k,D}^{\tiny{\mbox{b.h}}}(r)=M^{\frac{k-1}{2}}\sqrt{\frac{4\,[(D-1)k-(D-2)]}{(k-1)(D-2)}}r^{\frac{(k-1)(2-D)}{2k}}.\nonumber
\end{eqnarray}
Many comments can be made concerning this solution. Firstly, this
black hole solution depends on a unique integration constant $M$,
and for even $k$ the scalar field is always real provided that $M$
is positive constant while for odd $k$, the constant $M$ can be
positive or negative. The scalar field is well defined at the
horizon and blows-up at the singularity $r=0$. The solution given by
(\ref{xikbh}-\ref{potbh}-\ref{solbhkd}) reduces to the one derived
in the Einstein-Gauss-Bonnet case for $k=2$ \cite{Gaete:2013ixa}. It
is interesting to note that the standard GR-limit $k=1$ (which
yields to Einstein gravity with a negative cosmological constant) is
possible only in the zero-mass limit $M=0$ as it can be seen from
the expression of the scalar field. This is not surprising since for
$k=1$, the nonminimal coupling parameter (\ref{xikbh}) as well as
the potential (\ref{potbh}) vanish, and in this case, no-hair
theorems forbid the existence of black hole solutions. We will come
in details to this point in the Appendix.  We would like also to
emphasize that the allowed value of the nonminimal coupling
parameter (\ref{xikbh}) which depends on $(k,D)$ is bounded as
$\xi_{k,D}^{\tiny{\mbox{b.h}}}<1/4$ and its limit as $k$ goes to
infinity yields to the conformal coupling in $D$ dimensions,
$\lim_{k\to\infty}\xi_{k,D}^{\tiny{\mbox{b.h}}}=\xi_D$. Finally, we
may observe that in even dimension given by $D=2(k+1)$ which
corresponds to the Born-Infeld case, the lapse function
$F_{k,D}^{\tiny{\mbox{b.h}}}(r)$ has a BTZ-like form
\cite{Banados:1992wn}  as it occurs in the vacuum case with a base
manifold chosen to be non-Einstein \cite{Canfora:2010rh}.

\subsection{Planar AdS black hole stealth solutions}

The black hole stealth solution for $k\geq 2$ can be obtained in
analogue way that the one obtained in the Einstein-Gauss-Bonnet case
$k=2$ \cite{Gaete:2013ixa}; this means by combining the pure gravity
solution with planar base manifold \cite{Crisostomo:2000bb,
Aros:2000ij} with the stealth configuration. By stealth
configuration, we mean a non-trivial solution (with a non constant
scalar field) of the stealth equations
\begin{eqnarray}
{\cal G}^{(k)}_{\mu\nu}=0=T_{\mu\nu}, \label{321}
\end{eqnarray}
where both side (gravity and matter part) vanishes.

In fact, it is not difficult to show that a self-interacting
nonminimal scalar field   given by
\begin{eqnarray}
\label{phiste} \Phi(r)=A\,r^{\frac{2\xi}{4\xi-1}},
\end{eqnarray}
has a vanishing energy-momentum tensor (\ref{tmunusf})
$T_{\mu\nu}=0$ on the following $\xi-$dependent spacetime geometry
\begin{eqnarray} \label{geostel}
&&ds^{2}=-\Big(r^2-\frac{M}{r^{\frac{4(D-2)\xi-(D-3)}{4\xi-1}}}\Big)dt^{2}+\frac{dr^2}
{\Big(r^2-\frac{M}{r^{\frac{4(D-2)\xi-(D-3)}{4\xi-1}}}\Big)}+r^2d\vec{x}_{D-2}^2.
\end{eqnarray}
On the other hand, the pure gravity equations ${\cal
G}^{(k)}_{\mu\nu}=0$ has a black hole solution with a lapse function
$F(r)$ given by \cite{Crisostomo:2000bb,Aros:2000ij}
$$
F(r)=r^2-\frac{M}{r^{\frac{D-(2k+1)}{k}}}.
$$
Now, in order for this metric function to coincide with the the
stealth metric (\ref{geostel}), the nonminimal coupling parameter
must be chosen as
\begin{eqnarray}
\label{xikst}
&&\xi_{k,D}^{\tiny{\mbox{stealth}}}=\frac{(D-1)(k-1)}{4\left[Dk-(D-1)\right]},
\end{eqnarray}
and hence the mass term potential (\ref{potstealth}) becomes
\begin{eqnarray}
\label{ukds}
U_{k,D}^{\tiny{\mbox{stealth}}}(\Phi)=\frac{(k-1)(D-1)^2(D-1-k)}{8k^2\left[Dk-(D-1)\right]}\Phi^2.
\end{eqnarray}
Consequently, a topological black hole stealth solution of the
stealth equation (\ref{321}) is given by
\begin{eqnarray}
\label{ss}
&&ds^{2}=-F_{k,D}^{\tiny{\mbox{stealth}}}(r)dt^{2}+\frac{dr^2}{F_{k,D}^{\tiny{\mbox{stealth}}}(r)}+{r^{2}}d\vec{x}_{D-2}^2,\nonumber\\
&& F_{k,D}^{\tiny{\mbox{stealth}}}(r)=r^2-\frac{M}{r^{\frac{D-(2k+1)}{k}}},\\
\nonumber\\
&&
\Phi_{k,D}^{\tiny{\mbox{stealth}}}(r)=Ar^{\frac{(k-1)(1-D)}{2k}}.\nonumber
\end{eqnarray}
We may note that in contrast with the previous solution, the black
hole stealth solution  depends on two integration constants $M$ and
$A$, and in the vanishing $M$ limit, the solution reduces to a
stealth solution on the pure AdS background  \cite{ABMTZ}. The GR limit $k=1$ is also well defined
yielding to a metric that is noting but the topological
Schwarzschild-AdS-Tangherlini spacetime. This is not surprising
since in the GR limit case $k=1$, the nonminimal coupling parameter
as well as the potential vanish while the scalar field becomes
constant, and hence the energy-momentum tensor vanishes
 $T_{\mu\nu}=0$. In other words, in the GR-limit, the
stealth equations (\ref{321}) are equivalent to the pure Einstein
equations $G_{\mu\nu}-\frac{(D-1)(D-2)}{2} g_{\mu\nu}=0$. This class
of solutions is of particular interest since, up to now, the only
black hole stealth solution was the one obtained in
\cite{AyonBeato:2004ig} in the three-dimensional GR case with a
static BTZ metric \cite{Banados:1992wn}.

Let us go back to the $\xi-$dependent geometry (\ref{geostel})
allowing the existence of solution of $T_{\mu\nu}=0$. It is clear
from the expression of the metric that for a nonminimal coupling
parameter $\xi\in [\xi_{D+1},\frac{1}{4}[$ where
$\xi_{D+1}=(D-1)/(4D)$ is the conformal coupling in $(D+1)$
dimension, the asymptotic behavior of the metric as $r\to\infty$ is
faster than the usual AdS one. However, requiring the metric to be
also solution of the gravity part ${\cal G}^{(k)}_{\mu\nu}=0$, we
have seen that the parameter $\xi$ must take the form (\ref{xikst}),
and it is not difficult to prove that
$\xi_{k,D}^{\tiny{\mbox{stealth}}}<\xi_{D+1}$ while the critical
value $\xi_{D+1}$ is only reached at the limit $k\to\infty$. This
limit is also intriguing with respect to the effective mass square
(\ref{effmass}). The stealth solutions are not of constant scalar
curvature except for $M=0$, and in this case, the effective mass
(\ref{effmass}) becomes
$$
m_{\mbox{\tiny{eff}}}^2=-\frac{(D-1)^2 (k^2-1)}{4 k^2},
$$
and curiously enough this latter tends to the Breitenlohner-Freedman
bound $m^2_{BF}$ as $k$ goes to infinity,
$\lim_{k\to\infty}m_{\mbox{\tiny{eff}}}^2=m^2_{BF}$. To conclude, we
would like to point out  a certain symmetry between the black hole
and stealth solution as reflected by the following relations
\begin{subequations}
\begin{eqnarray*}
&&\xi_{k,D}^{\tiny{\mbox{stealth}}}=\xi_{k,D+1}^{\tiny{\mbox{b.h}}},\quad
F_{k,D}^{\tiny{\mbox{stealth}}}=F_{k,D+1}^{\tiny{\mbox{b.h}}},\quad
U_{k,D}^{\tiny{\mbox{stealth}}}= U_{k,D+1}^{\tiny{\mbox{b.h}}},\quad
\Phi_{k,D}^{\tiny{\mbox{stealth}}}\propto
\Phi_{k,D+1}^{\tiny{\mbox{b.h}}}.
\end{eqnarray*}
\end{subequations}
These relations can also be interpreted as follows: the two-parametric (with parameters $M$ and $A$) stealth solution in $D$ dimension given by (\ref{ss})
 can be promoted to the uniparametric (with constant $M$)
black hole  solution  in $(D+1)$  dimension (\ref{solbhkd})
through a {\it Kaluza-Klein oxidation}.

\section{Adding exact $(D-1)-$forms}
In the previous section, we have constructed two classes of
topological black hole solutions for a self-interacting nonminimal
scalar field with a gravity theory given by a particular Lovelock
action. The base manifold of these solutions is planar and these
configurations require a particular value of the nonminimal coupling
parameter (\ref{xikbh}-\ref{xikst}) which depends on the dimension
$D$ and the gravity theory $k\geq 2$. As it will be shown in the
Appendix, the existence of these solutions is strongly inherent to
the presence of the higher-order curvature terms of the Lovelock
theory. Indeed in the standard GR case $k=1$, we will establish that
black hole solutions with planar base manifold for a scalar field
nonminimally coupled with a possible mass term potential are only
possible in three dimensions yielding to the Martinez-Zanelli
solution \cite{Martinez:1996gn} \footnote{We will obtain this result
by considering an Ansatz for the metric that depends on a unique
lapse function.}. In the standard GR case, it has been shown
recently that the inclusion of multiple exact $p-$forms homogenously
distributed permits the construction of black holes with planar
horizon \cite{Bardoux:2012aw, Bardoux:2012tr} without any
restrictions on the dimension or on the value of the nonminimal
parameter \cite{Caldarelli:2013gqa}. Since, we are interested on
such solutions, we now propose to introduce appropriately some exact
$p-$forms in order to obtain topological black hole solutions with
arbitrary nonminimal coupling parameter. More precisely, we consider
the following action in arbitrary $D$ dimension
\begin{eqnarray}
{\cal S}_k=-\frac{1}{2k(D-3)!}&&\int
\sum_{p=0}^{k}\frac{C^k_p}{(D-2p)}~ L^{(p)}-\int d^{D}x\,\sqrt{-g}
\Big[ \frac{1}{2}\,\partial_{\mu}\Phi\partial^{\mu}\Phi+\frac{\xi}2
R\Phi^2 +U(\Phi)\Big]\nonumber\\
&&-\int\!
d^Dx\,\sqrt{-g}\left[\frac{\epsilon(\Phi)}{2(D-1)!}\sum_{i=1}^{D-2}
{\cal H}^{(i)}_{\alpha_1\cdots \alpha_{D-1}}{\cal
H}^{(i)\alpha_1\cdots \alpha_{D-1}}\right],\nonumber
\end{eqnarray}
where we have introduced $(D-2)$- fields ${\cal H}^{(i)}$ which are
exact $(D-1)$-forms, and where the potential is again the mass term
defined in (\ref{potstealth}). The coupling function between the
scalar field and the $(D-1)$-forms, $\epsilon(\Phi)$, depends on the
scalar field $\Phi$ as
\begin{eqnarray}\label{epsilon}
\epsilon(\Phi)=\sigma\,\Phi^{\frac{2(2-3k)\xi+k-1}{\xi(k-1)}}
\end{eqnarray}
where $\sigma$ is a coupling constant. We stress that the expression
of this coupling $\epsilon$ is not well-defined in the standard GR
case $k=1$. However, as mentioned before, the solutions in the
standard Einstein gravity have been obtained in
\cite{Caldarelli:2013gqa} for a more general class of potential than
the one considered here (\ref{potstealth}). Note that there exists
another particular value of the nonminimal coupling parameter
$\xi=\frac{k-1}{2(3k-2)}$ for which the coupling $\epsilon$ becomes
constant; we  will come to this point below.

The field equations obtained by varying the action with the
different dynamical fields $g_{\mu\nu}, \Phi$ and ${\cal H}^{(i)}$
read
\begin{subequations}
\label{eqsmotionkpf}
\begin{eqnarray}
&&{\cal G}^{(k)}_{\mu\nu}=T_{\mu\nu}+T_{\mu\nu}^{\tiny{extra}},
\qquad\qquad \partial_{\alpha}\left(\sqrt{-g}\,\epsilon(\Phi) {\cal H}^{(i)\alpha\alpha_1\cdots \alpha_{D-2}}\right)=0,\\
&&\Box \Phi = \xi R \Phi+\frac{d
U}{d\Phi}+\frac{1}{2}\frac{d\epsilon}{d
\Phi}\left[\sum_{i=1}^{D-2}\frac{1}{(D-1)!}{\cal
H}^{(i)}_{\alpha_1\cdots \alpha_{D-1}} {\cal H}^{(i)\alpha_1\cdots
\alpha_{D-1}}\right]=0,
\end{eqnarray}
\end{subequations}
where the extra piece in the energy-momentum tensor reads
\begin{eqnarray*}
\label{tmnextra}
T_{\mu\nu}^{\tiny{extra}}=\epsilon(\Phi)\,\sum_{i=1}^{D-2}\Big[\frac{1}{(D-2)!}
{\cal H}^{(i)}_{\mu\alpha_1\cdots \alpha_{D-2}} {\cal
H}_{\nu}^{(i)\alpha_1\cdots \alpha_{D-2}}-\frac{g_{\mu
\nu}}{2(D-1)!}{\cal H}^{(i)}_{\alpha_1\cdots \alpha_{D-1}} {\cal
H}^{(i)\alpha_1\cdots \alpha_{D-1}}\Big]
\end{eqnarray*}
Looking for a purely electrically homogenous Ansatz for the
$(D-1)-$forms as
\begin{eqnarray}
{\cal H}^{(i)}={\cal H}^{(i)}_{trx_1\cdots x_{i-1}x_{i+1}\cdots
x_{D-2}} (r) dt\wedge dr\wedge\cdots\wedge dx^{i-1}\wedge
dx^{i+1}\wedge \cdots \wedge dx^{D-2},
\end{eqnarray}
a solution of the field equations (\ref{eqsmotionkpf}) is given by
\begin{subequations}
\label{solps}
\begin{eqnarray}
&&ds^{2}=-\left(r^2-\frac{M}{r^{\frac{2\left[2(2k-1)\xi-(k-1)\right]}{(k-1)(1-4\xi)}}}\right)dt^{2}+
\frac{dr^2}{\left(r^2-\frac{M}{r^{\frac{2\left[2(2k-1)\xi-(k-1)\right]}{(k-1)(1-4\xi)}}}\right)}+{r^{2}}d\vec{x}_{D-2}^2,\\
\nonumber\\
&&\Phi(r)=\sqrt{\frac{M^{k-1}\left[  2\left( 3\,k-2 \right) \xi-k+1
\right]  \left( D-2 \right)
}{\big\{2\left[2(D-1)k-D\right]\xi-(k-1)(D-2)\big\} k\xi
}}\,r^{\frac{2\xi}{4\xi-1}},\\
\nonumber\\
&&{\cal H}^{(i)}=\frac{p}{\epsilon(\Phi)}r^{D-4}dt\wedge
dr\wedge\cdots\wedge dx^{i-1}\wedge dx^{i+1}\wedge \cdots \wedge
dx^{D-2},
\end{eqnarray}
\end{subequations}
where the constant $p$ is
defined by
\begin{eqnarray}
p={B\sqrt{\frac{-2\,\sigma\left(\xi-\xi_{k,D}^{\tiny{\mbox{stealth}}}\right)
\left(\xi-\xi_{k,D}^{\tiny{\mbox{b.h}}}\right)\Big[Dk-(D-1)\Big]\Big[(D-1)k-(D-2)\Big]}
{k-1}}},
\end{eqnarray}
and where the constants $\xi_{k,D}^{\tiny{\mbox{b.h}}}$ and
$\xi_{k,D}^{\tiny{\mbox{stealth}}}$ are the particular values of the
nonminimal parameter for which we have derived the previous
solutions in the pure scalar field case (\ref{xikbh}-\ref{xikst}).
In this last expression, the constant $B$ reads
\begin{eqnarray*}
B=\frac{4\,{\xi}^{{\frac { 2\left( 4\,k-3 \right) \xi-k+1}{4\xi\,
\left( k-1 \right) }}} \Big\{  \Big[
2\Big(2(D-1)k-D\Big)\xi-(k-1)(D-2)\Big] k \Big\} ^{{\frac { 2\left(
2\,k-1 \right) \xi-k+1}{4\xi\, \left( k-1 \right) }}}{M}^{{\frac {
\left( k-1 \right) \left( 1-4\,\xi \right) }{4\xi}}}}{ \left(
4\,\xi-1 \right) \Big\{ \Big[ 2\left( 3\,k-2 \right) \xi-k+1 \Big]
\left( D-2 \right) \Big\} ^{{\frac { 2\left( 3\,k-2 \right)
\xi-k+1}{4\xi\, \left( k-1 \right) }}}}
\end{eqnarray*}

As in the pure scalar field case, many comments can be made
concerning the solution obtained in the presence of these $(D-2)$
extra $(D-1)$-forms ${\cal H}^{(i)}$. Firstly, it is simple to see
that for $\xi=\xi_{k,D}^{\tiny{\mbox{stealth}}}$ or
$\xi=\xi_{k,D}^{\tiny{\mbox{b.h}}}$ the constant $p$ becomes zero
and the solutions are those found previously considering only a
scalar field nonminimally coupled with a mass term potential. From
the expression of the metric solution, we can see that for
$\xi>1/4$, the asymptotic behavior of the metric is faster than the
usual AdS one while for $\xi<1/4$, the dominant term as $r\to\infty$
is given by $F(r)\sim r^2$. From the expression of the scalar field,
it is easy to see that for a constant $M>0$, the allowed values of
$\xi$ in order to deal with a real solution are
$$
\xi\in\,\,
]0,\frac{k-1}{2(3k-2)}]\,\cup\,]\xi_{\mbox{\tiny{critical}}},+\infty[,\qquad
\quad
\xi_{\mbox{\tiny{critical}}}:=\frac{(k-1)(D-2)}{2\big[2(D-1)k-D\big]},
$$
while for $M<0$, the ranges are
\begin{eqnarray*}
&&\xi\in\,\, ]0,\frac{k-1}{2(3k-2)}]\,\cup\,]\xi_{\mbox{\tiny{critical}}},+\infty[,\qquad \mbox{for odd}\,\,\, k,\\
&& \xi\in\,\, [\frac{k-1}{2(3k-2)},
\xi_{\mbox{\tiny{critical}}}[,\qquad\qquad \quad \mbox{for
even}\,\,\, k.
\end{eqnarray*}

Solutions of constant scalar curvature $R=-D(D-1)$ are obtained for
two values of the nonminimal parameter
$$
\xi=\frac{(k-1)D}{4(D(k-1)+1)},\qquad\qquad
\xi=\frac{(k-1)(D-1)}{4\left[(D-1)k-(D-2)\right]}.
$$
For this last value of the parameter $\xi$, the effective square
mass (\ref{effmass}) becomes
$$
m_{\mbox{\tiny{eff}}}^2=\frac{(k-1)(k-3)(D-1)^2}{4},
$$
and it is intriguing to note that it saturates the
Breitenlohner-Freedman bound for $k=2$ \cite{Gaete:2013ixa} while
for $k=3$, the solution becomes massless.

{\it Pure axionic solution}: In order to be complete, we may look
for pure axionic solutions. This means a solution of the field
equations without considering the contribution of the scalar field,
\begin{equation}
{\cal G}^{(k)}_{\mu\nu}=T_{\mu\nu}^{\tiny{extra}}.
\end{equation}
Considering an Ansatz for the metric involving a unique metric
function, the integration of the field equations yields
\begin{subequations}
\begin{eqnarray}
&&ds^{2}=-\left(r^2-{M}{r^{\frac{2(k-1)}{k}}}\right)dt^{2}+
\frac{dr^2}{\left(r^2-{M}{r^{\frac{2(k-1)}{k}}}\right)}+{r^{2}}d\vec{x}_{D-2}^2,\\
\nonumber\\
&&{\cal H}^{(i)}= -M^{\frac{k}{2}}\,\sqrt{\frac{D-3}{k
\sigma}}\,r^{D-4}dt\wedge dr\wedge\cdots\wedge dx^{i-1}\wedge
dx^{i+1}\wedge \cdots \wedge dx^{D-2}.
\end{eqnarray}
\end{subequations}
We note that this solution can be obtained from  the solutions with
scalar field (\ref{solps})  by taking the well-defined limit
$\xi=\frac{k-1}{2(3k-2)}$. This is not surprising owing to the
choice of our coupling function $\epsilon$ defined in
(\ref{epsilon}) which becomes constant for
$\xi=\frac{k-1}{2(3k-2)}$. It is also interesting to note that in
this case, since the contribution of the scalar field is not longer
present, the GR limit $k=1$ is well-defined yielding a metric
function of the BTZ form \cite{Bardoux:2012aw}.

\section{Comments and conclusions}
Here, we have considered a gravity theory given by a particular
Lovelock Lagrangian labeled by an integer $k$ for which the
coefficients are fixed in order to have an unique AdS vacuum with a
fixed value of the cosmological constant. The matter part of our
action is concerned with a self-interacting scalar field
nonminimally coupled with a potential given by a mass term. For this
model labeled by the dimension $D$ and the integer $k$, we have
derived two classes of black hole solutions with  planar event
horizon topology for particular values
of the nonminimal coupling parameter depending on $D$ and $k$. The
first class of solutions is uniparametric and reduces to the pure
AdS metric without scalar field in the vanishing limit of the
parameter. The second class of solutions depends on two parameters
and is interpreted as a black hole stealth configuration. To be more
precise, we have shown the existence of a nontrivial
self-interacting scalar field with a vanishing energy-momentum
tensor with a  black hole metric solving the pure gravity equations.
In the last section, we have added to the starting action exact
$(D-1)-$forms minimally coupled to the scalar field. In this case
and for an appropriate coupling, we have been able to construct more
general black hole solutions with planar event horizon topology. All
these solutions generalize for an arbitrary $k\geq 2$ those obtained
in \cite{Gaete:2013ixa} in the Einstein-Gauss-Bonnet case $k=2$. In
the Appendix, we have established that these solutions may be viewed
as originated from a stealth configuration on a pure AdS background
through a Kerr-Schild transformation. We have also shown that their
standard GR counterpart $k=1$ can not be obtained along the same
lines, and hence the occurrence of such solutions is strongly
inherent to the presence of the higher-order curvature terms of the
Lovelock gravity theory. It seems then that the emergence of these
black hole solutions with planar event horizon topology is a
consequence of the higher-order curvature terms combined with the
existence of a stealth configuration on the pure AdS metric. Indeed,
as shown in \cite{ABMTZ}, static stealth configurations given by a
scalar field nonminimally coupled require the base manifold to be
planar. Indeed, stealth solutions with spherical or hyperboloid base
manifold are possible only in the non static case \cite{ABMTZ}. It
will be interesting to see whether these non static stealth
configurations with spherical or hyperboloid base manifold  can be
promoted as black hole solutions through a similar Kerr-Schild
transformation. In this case, since the scalar field depends
explicitly on the time as well as on the radial coordinates, the
metric function generated through the Kerr-Schild transformation
must probably depend also on these two coordinates. This will
considerably complicate the task of integrating the metric function.

In this paper, we have also derived a class of black hole stealth
configuration whose metric is a black hole solution of the pure
gravity equations \cite{Crisostomo:2000bb, Aros:2000ij}. These
metrics can be promoted to electrically charged black hole solutions
with a standard Maxwell source \cite{Crisostomo:2000bb,
Aros:2000ij}. It is then natural to ask whether one can derive the
electrically charged version of the solutions found here. As a first
task, it will be useful to derive, if possible, the electrically
charged version of the stealth configuration found in \cite{ABMTZ}.

We have seen that the existence of these solutions is strongly
inherent to the higher-order curvature terms of the Lovelock
theories. As it is well-known, the field equations associated to the
Lovelock gravity are of second order in spite of the presence of
these terms. In Ref. \cite{Oliva:2010eb}, a cubic gravity theory has
been constructed in five dimensions by requiring the trace of the
field equations to be proportional to the Lagrangian which in turn
implies that for an Ansatz metric of the "spherical" form, the field
equations are of second-order. It will be interesting to explore if
this five-dimensional cubic gravity theory and its generalizations
to higher odd dimension can accommodate the classes of solutions
found here.

In Ref. \cite{Crisostomo:2000bb, Aros:2000ij}, the authors have done
a complete study of  the thermodynamical properties of the pure
gravity solutions. The black hole stealth solutions obtained here
have the same lapse metric with a nontrivial matter source, and
hence it will be more than interesting to investigate the effects on
the thermodynamical quantities of the presence of this nontrivial
source. Note that the thermodynamics of general Lovelock gravity has
been analyzed in \cite{Cai:2003kt}.

\begin{acknowledgments}
We thank Julio Oliva and Sourya Ray for useful discussions. MB is
supported by BECA DOCTORAL CONICYT 21120271. MH was partially
supported by grant 1130423 from FONDECYT, by grant ACT 56 from
CONICYT and from CONICYT, Departamento de Relaciones Internacionales
``Programa Regional MATHAMSUD 13 MATH-05''.
\end{acknowledgments}

\section{Appendix}

\subsection{Stealth origin of the solutions}
In this Appendix, we will show that the AdS black hole solutions
obtained here can be viewed as originated from a stealth
configuration defined on the pure AdS metric through a Kerr-Schild
transformation.

Let us start with a self-interacting scalar field $\Phi$
nonminimally coupled whose stress tensor $T_{\mu\nu}$ is given by
\begin{eqnarray}
\label{tmn} T_{\mu \nu}=\partial_{\mu}\Phi\partial_{\nu} \Phi-g_{\mu
\nu}\Big(\frac{1}{2}\,\partial_{\sigma}\Phi\partial^{\sigma}\Phi+U(\Phi)\Big)+\xi\left(g_{\mu
\nu}\Box-\nabla_{\mu}\nabla_{\nu}+G_{\mu\nu}\right)\Phi^{2}.
\end{eqnarray}
As shown in Ref. \cite{ABMTZ}, a solution of the equation
$T_{\mu\nu}=0$ on the pure AdS metric
\begin{eqnarray}
ds^2=-r^2dt^2+\frac{dr^2}{r^2}+r^2d\vec{x}_{D-2}^2 \label{adsmetric}
\end{eqnarray}
is given by the following configuration
\begin{subequations}
\label{potg}
\begin{eqnarray}
\label{potg2}
U(\Phi)&=\frac{\xi}{{(1-4\xi)^2}}\,\left[2\,\xi\,b^{2}\Phi^{\frac{1-2\xi}{\xi}}-8\,(D-1)\left(\xi-\xi_{D}\right)
\left(2\,\xi\,b\,\Phi^{\frac{1}{2\xi}}-D \left(\xi-\xi_{D+1}\right)\,\Phi^{2} \right)\right],\\
\label{sf2} &\Phi(r)=(Ar+b)^{\frac{2\xi}{4\xi-1}}.
\end{eqnarray}
\end{subequations}
For technical reasons, we will restrict ourselves to the case $b=0$
which in turn implies that the stealth potential (\ref{potg2})
reduces to the mass term considered in this paper
(\ref{potstealth}). Let us first operate a Kerr-Schild
transformation on the AdS metric (\ref{adsmetric}) with a null and
geodesic vector $l=dt-\frac{dr}{r^2}$, and this without affecting
the scalar field. The transformed metric becomes after redefining
the time coordinate
\begin{eqnarray}
ds^2=-r^2\Big(1-f(r)\Big)dt^2+\frac{dr^2}{r^2\Big(1-f(r)\Big)}+r^2d\vec{x}_{D-2}^2.
\label{BAds}
\end{eqnarray}
It is easy to see that the components on-shell\footnote{By on-shell,
we mean using the expression of the potential (\ref{potg}), scalar
field (\ref{sf2}) and the Ansatz metric (\ref{BAds}) with $b=0$.} of
the energy-momentum tensor (\ref{tmn}) and the gravity tensor
satisfy the following identities
\begin{subequations}
\label{rel}
\begin{align}
{\cal G}_{\quad t}^{(k)t}&={\cal G}_{\quad r}^{(k)r},\qquad {\cal
G}_{\quad i}^{(k)i}=\frac{1}{(D-2)}\Big[r
\left({\cal G}_{\quad t}^{(k)t}\right)^{\prime}+{\cal G}_{\quad t}^{(k)t}(D-2)\Big]\\
T_t^t&=T_r^r,\qquad
T_i^i=\frac{(4\xi-1)}{4\xi(D-1)-(D-2)}\Big[r\left(T_t^t\right)^{\prime}+T_t^t(D-2)\Big]
\end{align}
\end{subequations}
Because of these relations (\ref{rel}), {\it a necessary condition}
for the field equations ${\cal G}^{(k)}_{\mu\nu}=T_{\mu\nu}$ to be
satisfied is that $\xi=0$ or $T_i^i=0={\cal G}_{\quad i}^{(k)i}$.
The condition  $T_i^i=0$ yields a second-order Cauchy equation for
the metric function $f$ whose solution reads
\begin{eqnarray}\label{fr}
f(r)=\frac{M_{1}}{r^{\frac{4(D-1)\xi-(D-2)}{4\xi-1}}}+\frac{M_{2}}{r^{\frac{(4\xi-1)D+1}{4\xi-1}}},
\end{eqnarray}
where $M_{1}$ and $M_{2}$ are two integration constants. Injecting
this metric function (\ref{fr}) into the condition ${\cal G}_{\quad
i}^{(k)i}=0$ yields
\begin{eqnarray}
\label{consg}
M_{1}^{2}\,\Big(\xi-\xi^{\tiny{(1)}}\Big)\Big(\xi-\xi_{k,D}^{\tiny{\mbox{b.h}}}\Big)
r^{2}+\beta_{2}M_{1}M_{2}\,\Big(\xi-\xi_{k,D}^{\tiny{\mbox{stealth}}}\Big)\Big(\xi-\xi_{k,D}^{\tiny{\mbox{b.h}}}\Big)
r\nonumber\\
+\beta_{3}\,M_{2}^{2}\,\Big(\xi-\xi_{k,D}^{\tiny{\mbox{stealth}}}\Big)\Big(\xi-\xi^{\tiny{(2)}}\Big)=0
\end{eqnarray}
where the  $\beta_{i}$ are non-vanishing constants and where we have
defined
\begin{eqnarray}
\xi^{\tiny{(1)}}=\frac{1}{4}\left[\frac{(k-1)D-2k+1}{(k-1)(D-1)}\right],\qquad
\xi^{\tiny{(2)}}=\frac{1}{4}\left[\frac{(k-1)D-k+2}{D(k-1)+2}\right].
\nonumber
\end{eqnarray}
As it can seen from $\xi^{\tiny{(1)}}$, these relations are valid
only for $k\geq2$. In fact, apart from the trivial solution
$M_{1}=M_{2}=0$ that yields to the pure AdS metric, there exists
{\it a priori} four options to solve the previous constraint
(\ref{consg})
\begin{eqnarray*}
&&\mbox{Option I}: \left\{M_{1}=0,\,
\xi=\xi_{k,D}^{\tiny{\mbox{stealth}}}\right\}
,\quad \mbox{Option II}: \left\{M_{1}=0,\,\xi= \xi^{\tiny{(2)}}\right\},\\
&& \mbox{Option III}: \left\{M_{2}=0,\,
\xi=\xi_{k,D}^{\tiny{\mbox{b.h}}}\right\}, \quad \mbox{Option IV}:
\left\{M_{2}=0,\,\xi= \xi^{(1)}\right\}.
\end{eqnarray*}
The options I and IV give rise to a metric function given by the
stealth metric $F_{k,D}^{\tiny{\mbox{stealth}}}$ while for the
options II and III, the metric becomes
$F_{k,D}^{\tiny{\mbox{b.h}}}$. However, it remains one independent
Einstein equation to be satisfied, that is ${\cal G}_{\quad
t}^{(k)t}=T_{\quad t}^{(k)t}$. In doing so, the options I and III
precisely yield to the two classes of solutions derived in this
paper. For the options II, the solution reduces to the stealth
configuration on the pure AdS metric \cite{ABMTZ}, and for the
option IV besides to the stealth configuration, there exists the
possibility with a vanishing scalar field defined on the stealth
metric $F_{k,D}^{\tiny{\mbox{stealth}}}$.

\subsection{Particular case of Einstein gravity $k=1$}

We will now consider the standard GR case $k=1$, and we will
establish that the unique black hole solution with planar base
manifold (with a unique metric function) of a scalar field
nonminimally coupled with a possible mass term potential  is the
Martinez-Zanelli solution in $D=3$
\cite{Martinez:1996gn}\footnote{We also assume that the scalar field
as well as the metric function only depends on the  coordinate
$r$.}. Hence, we consider the  following action
\begin{eqnarray}\label{action1}
S&=&\int d^{D}x\,\sqrt{-g}\,
\Big[\frac{1}{2}\left(R-2\Lambda\right)\Big]-\int
d^{D}x\,\sqrt{-g}\,
\Big(\frac{1}2\partial_{\mu}\Phi\partial^{\mu}\Phi+\frac{\xi}2
R\Phi^2 +\alpha \Phi^{2}\Big),
\end{eqnarray}
where $\Lambda=-\frac{1}{2}(D-1)(D-2)$ is the cosmological constant
and the potential is given by a mass term $U(\Phi)=\alpha \Phi^{2}$
where $\alpha$ is a constant. The field equations of (\ref{action1})
obtained by varying the action with respect to the different
dynamical fields read
\begin{eqnarray*}
E_{\mu \nu}&:=&G_{\mu\nu}+\Lambda
g_{\mu\nu}-\partial_{\mu}\Phi\partial_{\nu}
\Phi+g_{\mu\nu}\Big(\frac{1}{2}\,\partial_{\sigma}\Phi\partial^{\sigma}\Phi+\alpha
\Phi^{2}\Big)
-\xi\left(g_{\mu\nu}\Box-\nabla_{\mu}\nabla_{\nu}+G_{\mu\nu}\right)\Phi^{2}=0,\label{einseq1}
\\
\Box \Phi &=& \xi R \Phi+\frac{d U}{d\Phi}.
\end{eqnarray*}
We look for an Ansatz metric of the form
\begin{eqnarray}
ds^2=-F(r)\, dt^2+\frac{dr^2}{F(r)}+r^2d\vec{x}_{D-2}^2,
\end{eqnarray}
while the scalar field is assumed to depend only on the radial
coordinates, $\Phi=\Phi(r)$. The combination of the Einstein
equations combination $E_{t}^{t}-E_{r}^{r}=0$ implies that the
scalar field must be given by
\begin{equation}\label{scalarfield}
\Phi(r)=\frac{A}{\left(r+B\right)^{\frac{2\xi}{1-4\xi}}}
\end{equation}
where $A$ and $B$ are two integration constants. Substituting this
expression (\ref{scalarfield}) into the equation  $E_{t}^{t}=0$ (or
equivalently $E_{r}^{r}=0$), the metric function is obtained as
\begin{eqnarray}\label{metricfunction}
F(r)=\frac { 4\,\left( r+B \right) ^{1+\delta} \, \Big[\left( D-2
\right)  \left( 1+\delta \right) {r}^{2}-C{r}^{-D+3}-\alpha\,
h(r)\Big]}{4\, \left( 1+\delta \right)  \left( D-2 \right)  \left(
r+B \right) ^{1+\delta}+{A}^{2} \big[  \left( \delta-D+2 \right) r-B
\left( D-2 \right)  \big] \delta},
\end{eqnarray}
where for convenience, we have defined $\delta=4\xi/(1-4\xi)$ and
where $C$ is an integration constant. In this expression, the
function $h(r)$ can be given in an integral form as
\begin{eqnarray} h(r)&=&2\,{r}^{-D+3}{A}^{2} \left( 1+\delta
\right) \int \!{r}^{D-2} \left( r+B \right) ^{-\delta}{dr} \nonumber
\end{eqnarray}
or by a finite series
\begin{eqnarray*}
h(r)=2\,{r}^{-D+3}{A}^{2} \left( 1+\delta \right)\Big[\frac{r^{D-2}
(r+B)^{1-\delta}}{(1-\delta)}+\sum_{k=1}^{D-2}\frac{(-1)^{k}
(D-2)(D-3)\ldots (D-k-1)}{(1-\delta)(2-\delta)\ldots
(k+1-\delta)}\,\\ r^{D-k-2}\,\left(r+B\right)^{k+1-\delta} \Big],
\end{eqnarray*}
which is only valid for $\delta \neq 1,2,\ldots,(D-1)$.

The remaining of the analysis must be divided in two cases depending
if the coupling constant associated to the potential $\alpha$ is
chosen to be zero or not. For a non-vanishing $\alpha$, the
remaining independent Einstein equation $E_{i}^{i}=0$ will imply
that
\begin{equation}\label{alpha}
\alpha=\frac{\delta \left(\delta-D+2\right)
\left(\delta-D+1\right)}{8\left(\delta+1\right)},
\end{equation}
and $B=C=0$. This solution is nothing but a particular stealth
solution on the pure AdS metric.

We now turn to the case with a vanishing coupling constant
$\alpha=0$, and we split the analysis in two branches depending on
the sign of the constant $\delta$ (the option $\delta=0$ is
equivalent to $\xi=0$). In fact, for $\delta \neq 0$, we note that
substituting the metric function (\ref{metricfunction}) with
$\alpha=0$ into the equation $E_{i}^{i}=0$, we get the following
complicated expression after some algebraic manipulations
\begin{eqnarray}\label{eq1} (y-B)^{D} \left(\sum_{k=2}^{7}
\alpha_{k} y^{2\delta+k}+\sum_{k=1}^{7} \beta_{k}
y^{\delta+k}+\sum_{k=2}^{7} \gamma_{k}
y^{k}\right)+C\Big[\sum_{k=2}^{8} \epsilon_{k}
y^{2\delta+k}+\sum_{k=1}^{8} \eta_{k} y^{\delta+k}\nonumber\\
+A^{4} B \delta^{2} \left(-\delta+D-2\right) y^{2}
\left(-y+B\right)^{5}\Big]=0.
\end{eqnarray}
Here, we have defined $y=r+B$ and the different constants appearing
in this expression are not reported for simplicity.  For $\delta<0$,
the highest power of (\ref{eq1}) is ${D+7}$, and the vanishing of
the corresponding coefficient $\gamma_{7}$ implies that $A=0$ and
hence the scalar field $\Phi=0$. It is also easy to see that the
coefficient of $y^{2\delta+D+7}$ never vanishes  and finally we
conclude that solutions  do not exist  for $\delta<0$. Let us
now consider the case $\delta>0$ for which the highest power of
(\ref{eq1}) is $2\delta+D+7$ and its corresponding coefficient is
$\alpha_{7}$. In this situation, we have only two possible solutions
for a generic dimension $D$ which are $\delta=D-1$ or $\delta=D-2$.
For the first option  $\delta=D-1$,  the constant $B$ must vanish
and $C=-\frac{1}{4} A^{2} (D-1)$, yielding again to a particular
stealth configuration on the pure AdS metric. The remaining option
$\delta=D-2$ is the most interesting one. Indeed, the Eq.
(\ref{eq1}) becomes now
\begin{eqnarray}\label{eq2d} (y-B)^{D} \left(\sum_{k=2}^{5}
\alpha_{k,D}\, y^{2D+k-4}+\sum_{k=1}^{6} \beta_{k,D}\,
y^{D+k-2}+\sum_{k=2}^{5} \gamma_{k,D}\,
y^{k}\right)+\\C\Big(\sum_{k=2}^{8} \epsilon_{k,D}\,
y^{2D+k-4}+\sum_{k=1}^{7} \eta_{k,D}\, y^{D+k-2}\Big)=0.\nonumber
\end{eqnarray}
The highest possible powers are given by $y^{3D+1}$ and $y^{2D+4}$
and these latter coincide only in three dimension $D=3$. Otherwise
for $D>3$, the highest power is $y^{3D+1}$, and after some
computations we obtain that $B=C=0$ yielding again a particular
stealth configuration on the pure AdS metric. The case $D=3$ has
already been analyzed in details in \cite{AyonBeato:2001sb}, and the
only black hole solution is the Martinez-Zanelli solution
\cite{Martinez:1996gn}.


\end{document}